\documentclass[11pt]{article}
\usepackage{setspace}
\usepackage{epsfig}
\usepackage{palatino}
\usepackage{verbatim}
\usepackage{graphicx}
\usepackage{graphics}

\usepackage[round]{natbib}

\def\tr{{\mathrm T}}

\def\ol#1{{\overline{#1}}}
\def\eq{\!=\!}
\def\deg{$^{\circ}$}
\def\b{\begin{equation}}
\def\e{\end{equation}}
\def\ba{\begin{eqnarray}}
\def\ea{\end{eqnarray}}

\def\Emin{$E_{\mathrm{min}}$}
\def\sigmin{$\sigma_{\mathrm{min}}$}
\def\Epop{$E_{\mathrm{pop}}$}
\def\Erec{$E_{\mathrm{rec}}$}
\newcommand{\bm}[1]{\mbox{\boldmath $#1$}}

\begin{document}
\begin{center}
\begin{large}
{\bf WHEN RESPONSE VARIABILITY INCREASES  \\
\vspace*{0.05in} NEURAL NETWORK ROBUSTNESS TO SYNAPTIC NOISE\\}
\vspace*{0.2in}
\end{large}
\vspace*{0.3in}
\begin{large}
{\bf Gleb Basalyga and Emilio Salinas}
\vspace*{0.3in}
\end{large} \\
Department of Neurobiology and Anatomy \\
Wake Forest University School of Medicine \\
Winston-Salem, NC 27157-1010 \\
E-mail: gbasalyg@wfubmc.edu, esalinas@wfubmc.edu \\
 \vspace*{0.3in}
\today \\
 \vspace*{0.3in}
{\small Preliminary version of paper \\
to appear in
\textbf{\textit{Neural Computation}}}
 \vspace*{0.3in}
\end{center}

\centerline{\textbf{Abstract}}

\vspace{0.2in}

\noindent
Cortical sensory neurons are known to be highly variable, in the sense
that responses evoked by identical stimuli often change dramatically
from trial to trial. The origin of this variability is uncertain, but
it is usually interpreted as detrimental noise that reduces the
computational accuracy of neural circuits. Here we investigate the
possibility that such response variability might, in fact, be
beneficial, because it may partially compensate for a decrease in
accuracy due to stochastic changes in the synaptic strengths of a
network. We study the interplay between two kinds of noise, response
(or neuronal) noise and synaptic noise, by analyzing their joint
influence on the accuracy of neural networks trained to perform
various tasks. We find an interesting, generic interaction: when
fluctuations in the synaptic connections are proportional to their
strengths (multiplicative noise), a certain amount of response noise
in the input neurons can significantly improve network performance,
compared to the same network without response noise. Performance is
enhanced because response noise and multiplicative synaptic noise are
in some ways equivalent. So, if the algorithm used to find the optimal
synaptic weights can take into account the variability of the model
neurons, it can also take into account the variability of the
synapses. Thus, the connection patterns generated with response noise
are typically more resistant to synaptic degradation than those
obtained without response noise.  As a consequence of this interplay,
if multiplicative synaptic noise is present, it is better to have
response noise in the network than not to have it. These results are
demonstrated analytically for the most basic network consisting of two
input neurons and one output neuron performing a simple classification
task, but computer simulations show that the phenomenon persists in a
wide range of architectures, including recurrent (attractor) networks
and sensory-motor networks that perform coordinate transformations.
The results suggest that response variability could play an important
dynamic role in networks that continuously learn.

\newpage
\section{Introduction}

Neuronal networks face an inescapable tradeoff between learning new
associations and forgetting previously stored information. In
competitive learning models, this is sometimes referred to as the
stability-plasticity dilemma~\citep{carpenter87art2,hertz91b}: in
terms of inputs and outputs, learning to respond to new inputs will
interfere with the learned responses to familiar inputs. A
particularly severe form of performance degradation is known as
catastrophic interference~\citep{mccloskey89catastrophic}. It refers
to situations in which the learning of new information causes the
virtually complete loss of previously stored associations.

Biological networks must face a similar problem, because once a task
has been mastered, plasticity mechanisms will inevitably produce
further changes in the internal structural elements, leading to
decreased performance. That is, within sub-networks that have already
learned to perform a specific function, synaptic plasticity must at
least partly appear as a source of noise. In the cortex, this problem
must be quite significant, given that even primary sensory areas show a
large capacity for reorganization~\citep{XMSJ95,KM98,CLG01}. Some
mechanisms, such as homeostatic regulation~\citep{TN00} and specific
types of synaptic modification rules~\citep{HB04}, may help alleviate
the problem, but by and large, how nervous systems cope with it
remains unknown.

Another factor that is typically considered as a limitation for neural
computation capacity is response variability. The activity of cortical
neurons is highly variable, as measured either by the temporal
structure of spike trains produced during constant stimulation
conditions, or by spike counts collected in a given time interval and
compared across identical behavioral
trials~\citep{Dean81,nc:Softky+Koch:1992,SK93,HSKD96}. Some of the
biophysical factors that give rise to this variability, such as the
balance between excitation and inhibition, have been
identified~\citep{SK93,SN94,SZ98}. But its functional significance, if
any, is not understood.

Here we consider a possible relationship between the two sources of
randomness just discussed, whereby response variability helps
counteract the destabilizing effects of synaptic changes. Although
noise generally hampers performance, recent studies have shown that
in nonlinear dynamical systems such as neural networks this is not
always the case. The best known example is stochastic resonance, in
which noise enhances the sensitivity of sensory neurons to weak
periodic signals~\citep{LM96,Gammaitoni98SR,Nozaki99}, but noise may
play other constructive roles as well. For instance, when a system has
an internal source of noise, an externally added noise can reduce the
total noise of the output~\citep{Vilar-Rubi-2000}. Also, adding noise
to the synaptic connections of a network during learning produces
networks that, after training, are more robust to synaptic corruption
and have a higher capacity to generalize~\citep{murray94enhanced}.

In this paper we study another beneficial effect of noise on neural
network performance. In this case, adding randomness to the neural
responses reduces the impact of fluctuations in synaptic strength.
That is, here, performance depends on two sources of variability,
response noise and synaptic noise, and adding some amount of response
noise produces better performance than having synaptic noise alone.
The reason for this paradoxical effect is that response noise acts as
a regularization factor that favors connectivity matrices with many
small synaptic weights over connectivity matrices with few large
weights, and this minimizes the impact of a synapse that is lost or
has a wrong value. We study this regularization effect in three
different cases: (1) a classification task, which in its simplest
instantiation can be studied analytically, (2) a sensory-motor
transformation, and (3) an attractor network that produces
self-sustained activity. For the latter two, the interaction between
noise terms is demonstrated by extensive numerical simulations.

\section{General Framework}
\label{general}

First we consider networks with two layers, an input layer that
contains $N$ sensory neurons and an output layer with $K$ output
neurons. A matrix $\bm{r}$ is used to denote the firing rates of the
input neurons in response to $M$ stimuli, so $r_{ij}$ is the firing
rate of input unit $i$ when stimulus $j$ is presented.  These rates
have a mean component $\ol{\bm{r}}$ plus noise, as described in detail
below.  The output units are driven by the first layer responses, such
that the firing rate of output unit $k$ evoked by stimulus $j$ is
\b
    R_{kj} = \sum_{i=1}^N w_{ki} \, r_{ij} ,
    \label{Rdriv}
\e
or in matrix form, $\bm{R}= \bm{w} \bm{r}$, where $\bm{w}$ is the
$K\!\times\!N$ matrix of synaptic connections between input and output
neurons. The output neurons also have a set of desired responses
$\bm{F}$, where $F_{kj}$ is the firing rate that output unit $k$
should produce when stimulus $j$ is presented. In other words, $\bm F$
contains target values that the outputs are supposed to learn. The
error $E$ is the mean squared difference between the actual driven
responses $R_{kj}$ and the desired ones,
\b
    E = \left< \frac{1}{KM} \, \sum_{k=1}^K\sum_{j=1}^M
               \left( R_{kj} - F_{kj} \right)^2
        \right> ,
    \label{error}
\e
or in matrix notation,
\b
    E = \frac{1}{KM}
        \left< \mbox{Tr}
            \left[(\bm{w} \bm{r} - \bm{F})
                  (\bm{w} \bm{r} - \bm{F})^\tr
            \right]
        \right> .
    \label{errormatrix}
\e
Here, $\mbox{Tr}(\bm{A}) = \sum_i A_{ii}$ is the trace of a matrix and
the angle brackets indicate an average over multiple trials, which
corresponds to multiple samples of the noise in the inputs $\bm{r}$.
The optimal synaptic connections $\ol{\bm{W}}$ are those that make the
error as small as possible.  These can be found by computing the
derivative of Equation (\ref{errormatrix}) with respect to $\bm{w}$
(or with respect to $w_{ab}$, if the summations are written
explicitly) and setting the result equal to zero~\citep[see
e.g.,][]{GolubLoan96a}.  These steps give
\b
    \ol{\bm{W}} = \bm{F} \, \ol{\bm{r}}^\tr \bm{C}^{-1} ,
    \label{wopt}
\e
where $\ol{\bm{r}}\eq \left<\bm{r}\right>$ and $\bm{C}^{-1}$ is the
inverse (or the pseudo-inverse) of the correlation matrix
$\bm{C} = \left<\bm{r} \bm{r}^\tr\right>$.

The general outline of the computer experiments proceeds in five
steps as follows. First, the matrix $\ol{\bm r}$ with the mean input
responses is generated together with the desired output responses
$\bm{F}$. These two quantities define the input-output transformation
that the network is supposed to implement.  Second, response noise is
added to the mean input rates, such that
\b
    r_{ij} = \ol{r}_{ij} (1 + \eta_{ij}).
    \label{inputnoise1}
\e
The random variables $\eta_{ij}$ are independently drawn from a
distribution with zero mean and variance $\sigma_r^2$,
\ba
    \left< \eta_{ij} \right > & = & 0 \nonumber \\
    \left< \eta^2_{ij} \right> & = & \sigma_r^2 ,
    \label{inputnoise1a}
\ea
where the brackets again denote an average over trials. We refer to
this as multiplicative noise.  Third, the optimal connections are
found using Equation (\ref{wopt}). Note that these connections take
into account the response noise through its effect on the correlation
matrix $\bm{C}$. Fourth, the connections are corrupted by
multiplicative synaptic noise with variance $\sigma_W^2$, that is
\b
    W_{ij} = \ol{W}_{ij} (1 + \epsilon_{ij}),
    \label{weightnoisegeneral}
\e
where
\ba
    \left< \epsilon_{ij} \right> & = & 0 \nonumber \\
    \left< \epsilon^2_{ij} \right> & = & \sigma_W^2 .
\ea
Finally, the network's performance is evaluated. For this, we measure
the network error $E_W$, which is the square error obtained with the
optimal but corrupted weights $\bm{W}$, averaged over both types of
noise,
\b
    E_W = \frac{1}{KM}
          \left< \mbox{Tr} \left[
             (\bm{W} \bm{r} - \bm{F}) (\bm{W} \bm{r} - \bm{F})^\tr
          \right] \right> .
    \label{errornet}
\e
Thus, the brackets in this case indicate an average over multiple
trials and multiple networks, i.e., multiple corruptions of the
optimal weights $\ol{\bm{W}}$.

The main result we report here is an interaction between the two types
of noise: in all the network architectures that we have explored, for
a fixed amount of synaptic noise $\sigma_W$, the best performance is
typically found when the response noise has a certain nonzero
variance. So, given that there is synaptic noise in the network, it is
better to have some response noise rather than to have none.

Before addressing the first example, we should highlight some features
of the chosen noise models. Regarding response noise, Equations
(\ref{inputnoise1}, \ref{inputnoise1a}), other models were tested in
which the fluctuations were additive rather than multiplicative. Also,
Gaussian, uniform and exponential distributions were tested.  The
results for all combinations were qualitatively the same, so the shape
of the response noise distribution does not seem to play an important
role; what counts is mainly the variance.  On the other hand, the
benefit of response noise is observed only when the synaptic noise is
multiplicative; it disappears with additive synaptic noise.  However,
we do test several variants of the multiplicative model, including one
in which the random variables $\epsilon_{ij}$ are drawn from a
Gaussian distribution and another in which they are binary, 0 or -1.
The latter case represents a situation in which connections are
eliminated randomly with a fixed probability.

\section{Noise Interactions in a Classification Task}

First we consider a task in which the two-layer, fully connected
network is used to approximate a binary function. The task is to
classify $M$ stimuli on the basis of the $N$ input firing rates evoked
by each stimulus. Only one output neuron is needed, so $K\eq1$.  The
desired response of this output neuron is the classification function
\b
   F_j = \left\{\begin{array}{l}
            1 \ \: \: \mbox{if } j \leq M/2 \\
            0 \ \: \: \mbox{else} ,
         \end{array}\right.
   \label{desiredoutput}
\e
where $j$ goes from 1 to $M$. Therefore, the job of the output unit is
to produce a 1 for the first $M/2$ input stimuli and a 0 for the rest.

\subsection{A Minimal Network}

In order to obtain an analytical description of the noise
interactions, we first consider the simplest possible network that
exhibits the effect, which consists of two input neurons and two
stimuli. Thus, $N\eq M\eq 2$ and the desired output is
$\bm{F} = \left(1, 0\right)$. Note that, with a single output neuron,
the matrices $\bm{W}$ and $\bm{F}$ become row vectors. Now we proceed
according to the five steps outlined in the preceding section --- the
goal is to show analytically that, in the presence of synaptic noise,
performance is typically better for a nonzero amount of response
noise.

The matrix of mean input firing rates is set to
\b
   \ol{\bm{r}} = \left( \begin{array}{cc}
                   1 & r_0 \\
                   r_0 & 1 \\
                 \end{array} \right) ,
    \label{inputmeanmatrix}
\e
where $r_0$ is a parameter that controls the difficulty of the
classification. When it is close to 1, the pairs of responses evoked
by the two stimuli are very similar and large errors in the output are
expected; when it is close to 0, the input responses are most
different and the classification should be more accurate. After
combining the mean responses with multiplicative noise, as prescribed
by Equation (\ref{inputnoise1}), the input responses in a given trial
become
\b
   \bm{r} = \left( \begin{array}{cc}
              1 + \eta_{11}     &  r_0 (1+\eta_{12}) \\
              r_0 (1+\eta_{21}) &  1 + \eta_{22} \\
            \end{array} \right) .
   \label{multinputmeanmatrix}
\e
Assuming that the fluctuations are independent across neurons, the
correlation matrix is, therefore,
\b
   \bm{C} = \left< \bm{r} \bm{r}^\tr \right>
          = \left(\begin{array}{cc}
              (1+r_0^2)(1+\sigma_r^2)  &  2 r_0  \\
              2 r_0  &  (1+r_0^2)(1+\sigma_r^2)  \\
            \end{array} \right) .
   \label{multicorrelationmatrix}
\e
Next, after calculating the inverse of $\bm{C}$, Equation (\ref{wopt})
is used to find the optimal weights, which are
\ba
   \ol{W}_1 & = & \frac{\sigma_r^2 (1+r_0^2) + (1-r_0^2)}
                       {(1+\sigma_r^2)^2 \, (1+r_0^2)^2 - 4 r_0^2} \nonumber \\
   \ol{W}_2 & = & \frac{\sigma_r^2 (1+r_0^2) - (1-r_0^2)}
                       {(1+\sigma_r^2)^2 \, (1+r_0^2)^2 - 4 r_0^2} \: r_0 \, .
   \label{multiwopt1}
\ea
Notice that these connections take into account the response
variability through their dependence on $\sigma_r$. The next step is
to corrupt these synaptic weights as prescribed by Equation
(\ref{weightnoisegeneral}), and substitute the resulting expressions
into Equation (\ref{errornet}). After making all the substitutions,
calculating the averages and simplifying, we obtain the average error,
%\b
%   E_W = (1 + \sigma_W^2)
%            (\ol{W}^2_1 + \ol{W}^2_2) (1 + \sigma_r^2) (1 + r_0^2)
%         + 4 r_0 \ol{W}_1 \ol{W}_2
%         - 2 r_0 \ol{W}_2 - 2 \ol{W}_1 + 1 .
%\e
\b
   E_W = \frac{1}{2} \left(
            \sigma_W^2 (\ol{W}^2_1 + \ol{W}^2_2)
                       (1 + \sigma_r^2) (1 + r_0^2)
                          - \ol{W}_1 - r_0 \ol{W}_2 + 1
         \right) .
   \label{multierror2}
\e
This is the average square difference between the desired and actual
responses of the output neuron given the two types of noise. It is a
function only of three parameters, $\sigma_r$, $\sigma_W$ and $r_0$,
because the optimal weights themselves depend on $\sigma_r$ and $r_0$.

The interaction between noise terms for this simple $N\eq K\eq 2$ case
is illustrated in Fig.~1A, which plots the error as a function of
$\sigma_r$ with and without synaptic variability.  Here, dashed and
solid lines represent the theoretical results given by Equations
(\ref{multiwopt1}, \ref{multierror2}) and symbols correspond to
simulation results averaged over $1000$ networks and $100$ trials per
network. Without synaptic noise (dashed line), the error increases
monotonically with $\sigma_r$, as one would normally expect when
adding response variability. In contrast, when $\sigma_W\eq 0.15$, 0.2
or 0.25 (solid lines), the error initially decreases and then starts
increasing again, slowly approaching the curve obtained with response
noise alone.

\begin{figure*}[tb]
\centerline{\epsfig{figure=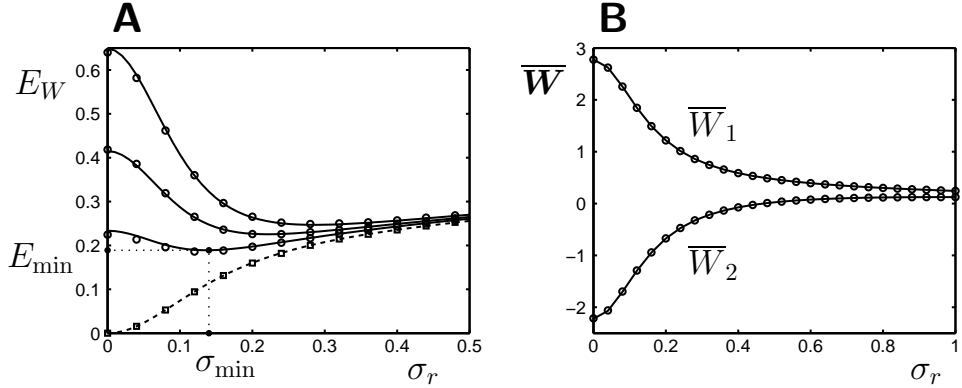,width=5.0in}}
\caption{\label{WeightSpace}
Noise interaction for a simple network of two input neurons and one
output neuron ($K\eq 1$, $N\eq M\eq 2$). Both input responses and
synaptic weights were corrupted by multiplicative Gaussian noise. For
all curves, solid lines are theoretical results and symbols are
simulation results averaged over $1000$ networks and $100$ trials per
network. In all cases, $r_0\eq 0.8$.
(A) Average square difference between observed and desired output
responses, $E_W$, as a function of the standard deviation (SD) of the
response noise, $\sigma_r$.  Squares and dashed line correspond to the
error without synaptic noise ($\sigma_{W}\eq 0$); circles and
continuous lines correspond to the error with synaptic noise
($\sigma_{W}\eq 0.15, 0.20, 0.25$).
(B) Dependence of the (uncorrupted) optimal weights $\ol{\bm{W}}$ on
$\sigma_r$. }
\end{figure*}

Figure 1B shows how the optimal weights depend on $\sigma_r$. The
solid lines were obtained from Equations (\ref{multiwopt1}) above.
The curves show that the effect of response noise is to decrease the
absolute values of the optimal synaptic weights.  Intuitively, that is
why response variability is advantageous; smaller synaptic weights
also mean smaller synaptic fluctuations, because their standard
deviation (SD) is proportional to the mean values.  So, there is a
tradeoff: the intrinsic effect of increasing $\sigma_r$ is to increase
the error, but with synaptic noise present, $\sigma_r$ also decreases
the magnitude of the weights, which lowers the impact of the synaptic
fluctuations. That the impact of synaptic noise grows directly with
the magnitude of the weights is also apparent from the first term in
Equation (\ref{multierror2}).

The magnitude of the noise interaction can be quantified by the
ratio \Emin$/E_0$, where the numerator is the minimal value of the
error curve and the denominator is the error obtained when only
synaptic noise is present, that is, when $\sigma_r\eq 0$.  The
minimum error \Emin\ occurs at the optimal value of $\sigma_r$,
denoted as \sigmin. The ratio \Emin$/E_0$ is equal to 1 if response
variability provides no advantage and approaches 0 as \sigmin\
cancels more of the error due to synaptic noise.  For the lowest
solid curve in Fig.~1A the ratio is approximately 0.8, so response
variability cancels about 20\% of the square error generated by
synaptic fluctuations. Note, however, that in these examples the
error is below $E_0$ for a large range of values of $\sigma_r$, not
only near \sigmin, so response noise may be beneficial even if it is
not precisely matched to the amount of synaptic noise.

\begin{figure*}[tb!]
\centerline{\epsfig{figure=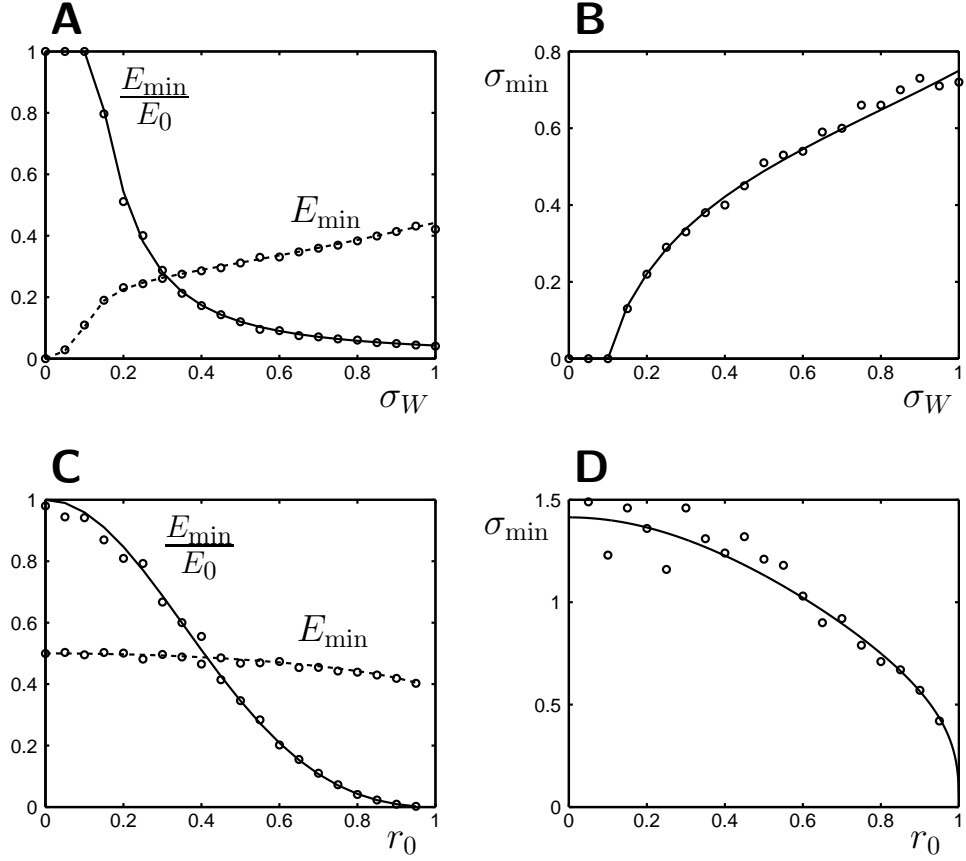,width=5.0in}}
\caption{\label{MultiNoise1}
Optimal amount of response noise in the minimal classification
network.  Same network with two sensory neurons and one output neuron
as in Fig.~1.  Lines and symbols indicate theoretical and simulation
results, respectively, averaged over $1000$ networks and $100$ trials
per network.
(A) Strength of the noise interaction quantified by \Emin\ (dashed
line) and \Emin$/E_0$ (solid line), as a function of $\sigma_{W}$, which
determines the synaptic variability.  Here and in B, $r_0\eq 0.8$.
(B) Optimal amount of response variability, \sigmin, as a function of
$\sigma_{W}$, for the same data in A\@.
(C) Strength of the noise interaction as a function of $r_0$, which
parameterizes the discriminability of the mean input responses evoked
by the two stimuli. Here and in D, $\sigma_W\eq 1$.
(D) \sigmin, as a function of $r_0$ for the same data in C\@.  }
\end{figure*}

Figure 2 further characterizes the strength of the interaction between
the two types of noise.  Figures 2A, B show how the error and the
optimal amount of response variability vary as functions of
$\sigma_W$. These graphs indicate that the fraction of the error that
$\sigma_r$ is able to compensate for, as well as the optimal amount of
response noise, increases with the SD of the synaptic noise.  The
minimum error, \Emin, grows steadily with $\sigma_W$ --- clearly,
$\sigma_r$ cannot completely compensate for synaptic corruption.
Also, $\sigma_W$ has to be bigger than a critical value for the noise
interaction to be observed ($\sigma_W\!>\!0.1$, approximately).
However, except when synaptic noise is very small, the optimal
strategy is to add some response noise to the network.

As in the previous figure, symbols and lines in Fig. 2 correspond to
simulation and theoretical results, respectively. To obtain the
latter, the key is to calculate \sigmin. This is done by, first,
substituting the optimal synaptic weights of Equation
(\ref{multiwopt1}) into the expression for the average error, Equation
(\ref{multierror2}), and second, calculating the derivative of the
error with respect to $\sigma_r^2$ and equating it to zero. The
resulting expression gives $\sigma^2_{\mathrm{min}}$ as a function of
the only two remaining parameters, $\sigma_W$ and $r_0$.  The
dependence, however, is highly nonlinear, so in general the solution
is implicit:
\ba
     \lefteqn{\sigma_r^8 \, (1 - \sigma_W^2) +
           2 \sigma_r^6 \, (1 + a^2(1 - 2 \sigma_W^2)) +
           6 \sigma_r^4 a^2 \, (1 - \sigma_W^2) + \mbox{} }
           \hspace*{1.2cm} \nonumber \\
     &  &  2 \sigma_r^2 a^2 \, (1 + a^2 + 2 a^2 \sigma_W^2 - 4 \sigma_W^2) +
           a^4 (1 + 3 \sigma_W^2) -
           4 a^2 \sigma_W^2 \:\: = \:\: 0 \, ,
     \label{implicit}
\ea
where
\b
   a \equiv \frac{1 - r_0^2}{1 + r_0^2} \, .
\e
The value of $\sigma_r$ that makes Equation (\ref{implicit}) true is
\sigmin. For Figs.~2A, B, the zero of the polynomial was found
numerically for each combination of $r_0$ and $\sigma_W$.

Figures 2C, D show how \Emin, \Emin/$E_0$ and \sigmin\ depend on the
separation between evoked input responses, as parameterized by $r_0$.
For these two plots, we chose a special case in which \sigmin\ can be
obtained analytically from Equation (\ref{implicit}): $\sigma_W\eq 1$.
In this particular case the dependence of \sigmin\ on $r_0$ has a
closed form,
\b
   \sigma_{\mathrm{min}}^2 = \frac{(1-r_0^2)^{2/3}}{1+r_0^2}
               \left( (1+r_0)^{2/3} + (1-r_0)^{2/3} \right) .
   \label{sigmamin}
\e
This function is shown in Fig.~2D.  In general, the numerical
simulations are in good agreement with the theory, except that the
scatter in Fig.~2D tends to increase as $r_0$ approaches 0. This is
due to a key feature of the noise interaction, which is that it
depends on the overlap between input responses across stimuli. This
can be seen as follows.

First, notice that in Fig.~2C the relative error approaches 1 as $r_0$
gets closer to 0. Thus, the noise interaction becomes weaker when
there is less overlap between input responses, which is precisely what
$r_0$ represents in Equation (\ref{inputmeanmatrix}). If there is no
overlap at all, the benefit of response noise vanishes. This fact
explains why more than one neuron is needed to observe the noise
interaction in the first place.  This observation can be demonstrated
analytically by setting $r_0\eq 0$ in Equations (\ref{multiwopt1}) and
(\ref{multierror2}), in which case the average square error becomes
\b
   E_W(r_0\eq 0) = \frac{1}{2} \left(
                     \frac{\sigma_W^2 - 1}{1 + \sigma_r^2} + 1
                   \right) .
   \label{r0error}
\e
This result has interesting implications.  If $\sigma_W^2\eq 1$,
response noise makes no difference, so there is no optimal value.  If
$\sigma_W^2\!<\!1$, the error increases monotonically with response
noise, so the optimal value is 0.  And if $\sigma_W^2\!>\!1$, the
optimal strategy is to add as much noise as possible! In this case,
the variance of the output neuron is so high that there is no hope of
finding a reasonable solution; the best thing to do is set the mean
weights to zero, disconnecting the output unit. Thus, without overlap,
either the synaptic noise is so high that the network is effectively
useless, or, if $\sigma_W$ is tolerable, response noise does not
improve performance. At $r_0\eq 0$, the numerical solutions oscillate
between these two extremes, producing an average error of 0.5
(leftmost point in Fig.~2C). In general, however, with non-zero
overlap there is a true optimal amount of response noise, and the more
overlap there is, the larger its benefit, as shown in Fig.~2C\@.

The simulation data points in Fig.~2 were obtained using fluctuations
$\epsilon$ and $\eta$ in Equations (\ref{weightnoisegeneral}) and
(\ref{multinputmeanmatrix}), respectively, sampled from Gaussian
distributions. The results, however, were virtually identical when the
distribution functions were either uniform or exponential.  Thus, as
noted earlier, the exact shapes of the noise distributions do not
restrict the observed effect.

\subsection{Regularization by Noise}
\label{RegSect}

Above, we mentioned that response noise tends to decrease the absolute
value of the optimal synaptic weights. Why is this? The reason is that
minimization of the mean square error in the presence of response
noise is mathematically equivalent to minimization of the same error
without response noise but with an imposed constraint forcing the
optimal weights to be small. This is as follows.

Consider Equation (\ref{wopt}), which specifies the optimal weights in
the two-layer network.  Response noise enters into the expression
through the correlation matrix. By separating the input responses into
mean plus noise, we have
\ba
   \bm{C} & = & \left< (\ol{\bm{r}} + \bm{\eta})
                       (\ol{\bm{r}} + \bm{\eta})^{\tr} \right>
                \nonumber \\
          & = & \ol{\bm{r}} \, \ol{\bm{r}}^{\tr} +
                \left< \bm{\eta} \bm{\eta}^{\tr} \right>
                \nonumber \\
          & = & \ol{\bm{r}} \, \ol{\bm{r}}^{\tr} +
                \bm{D}_{\!\sigma} \, ,
   \label{newcorr}
\ea
where we have assumed that the noise is additive and uncorrelated
across neurons (additivity is considered for simplicity but is not
necessary). This results in the diagonal matrix $\bm{D}_{\!\sigma}$
containing the variances of individual units, such that element $j$
along the diagonal is the total variance, summed over all stimuli, of
input neuron $j$. Thus, uncorrelated response noise adds a diagonal
matrix to the correlation between average responses.  In that case,
Equation (\ref{wopt}) can be rewritten as
\b
    \ol{\bm{W}} = \bm{F} \, \ol{\bm{r}}^\tr
                  \left( \ol{\bm{r}} \, \ol{\bm{r}}^{\tr}
                         + \bm{D}_{\!\sigma}
                  \right)^{-1} .
    \label{wopt1}
\e

Now consider the mean square error without any noise but with an
additional term that penalizes large weights. To restrict, for
instance, the total synaptic weight provided by each input neuron, add
the penalty term
\b
   \frac{1}{KM} \sum_{i, j} \lambda_i \, w_{ij}^2
   \label{wcost}
\e
to the original error expression, Equation (\ref{errormatrix}). Here,
$\lambda_i$ determines how much input neuron $i$ is taxed for its
total synaptic weight. Rewriting this as a trace, the total error to
be minimized in this case becomes
\b
    E = \frac{1}{KM} \left(
          \left< \mbox{Tr}
              \left[(\bm{w} \ol{\bm{r}} - \bm{F})
                    (\bm{w} \ol{\bm{r}} - \bm{F})^\tr
              \right]
          \right> +
          \mbox{Tr}\left(\bm{w}^{\tr} \bm{D}_{\!\lambda} \bm{w} \right)
        \right) .
\e
where $\bm{D}_{\!\lambda}$ is a diagonal matrix that contains the
penalty coefficients $\lambda_i$ along the diagonal. The synaptic
weights that minimize this error function are given by
\b
   \bm{F} \, \ol{\bm{r}}^\tr
      \left( \ol{\bm{r}} \, \ol{\bm{r}}^{\tr}
             + \bm{D}_{\!\lambda}
      \right)^{-1} \! .
   \label{wopt2}
\e
But this solution has exactly the same form as Equation (\ref{wopt1}),
which minimizes the error in the presence of response noise alone,
without any other constraints. Therefore, adding response noise is
equivalent to imposing a constraint on the magnitude of the synaptic
weights, with more noise corresponding to smaller weights. The penalty
term in Equation (\ref{wcost}) can also be interpreted as a
regularization term, which refers to a common type of constraint used
to force the solution of an optimization problem to vary
smoothly~\citep{Hint89,Hayk99}. Therefore, as has been pointed out
previously~\citep{Bish95}, the effect of response fluctuations can be
described as regularization by noise.

In our model, we assumed that the fluctuations in synaptic connections
are proportional to their size. What happens, then, is that response
noise forces the optimal weights to be small, and this significantly
decreases the part of the error that depends on $\sigma_W$.  In this
way, smaller synaptic weights --- and therefore a nonzero $\sigma_r$
--- typically lead to smaller output errors.

Another way to look at the relationship between the two types of noise is
to calculate the optimal mean synaptic weights taking the synaptic
variability directly into account. For simplicity, suppose that there
is no response noise. Substitute Equation (\ref{weightnoisegeneral})
directly into Equation (\ref{errormatrix}) and minimize with respect
to $\ol{\bm{W}}$, now averaging over the synaptic fluctuations. With
multiplicative noise the result is again an expression similar to
Equations (\ref{wopt1}) and (\ref{wopt2}), where a correction
proportional to the synaptic variance is added to the diagonal of the
correlation matrix.  In contrast, with additive synaptic noise the
resulting optimal weights are exactly the same as without any
variability, because this type of noise cannot be compensated for.
Therefore, the recipe for counteracting response noise is equivalent
to the recipe for counteracting multiplicative synaptic noise. An
argument outlining why this is generally true is presented in the
Discussion, Section~\ref{disc1}.

\subsection{Classification in Larger Networks}

When the simple classification task is extended to larger numbers of
first-layer neurons ($N\!>2$) and more input stimuli to classify
($M\!>2$), an important question can be studied: how does the
interaction between synaptic and response noise depend on the
dimensionality of the problem, that is, on $N$ and $M$?  To address
this issue we did the following. Each entry in the $N\times M$ matrix
$\ol{\bm{r}}$ of mean responses was taken from a uniform distribution
between 0 and 1.  The desired output still consisted of a single
neuron's response given by Equation (\ref{desiredoutput}), as before.
So, each one of the $M$ input stimuli evoked a set of $N$ neuronal
responses, each set drawn from the same distribution, and the output
neuron had to divide the $M$ evoked firing rate patterns into two
categories. The optimal amount of response noise was found, and the
process was repeated for different combinations of $N$ and $M$\@.

\begin{figure*}[tb!]
\centerline{\epsfig{figure=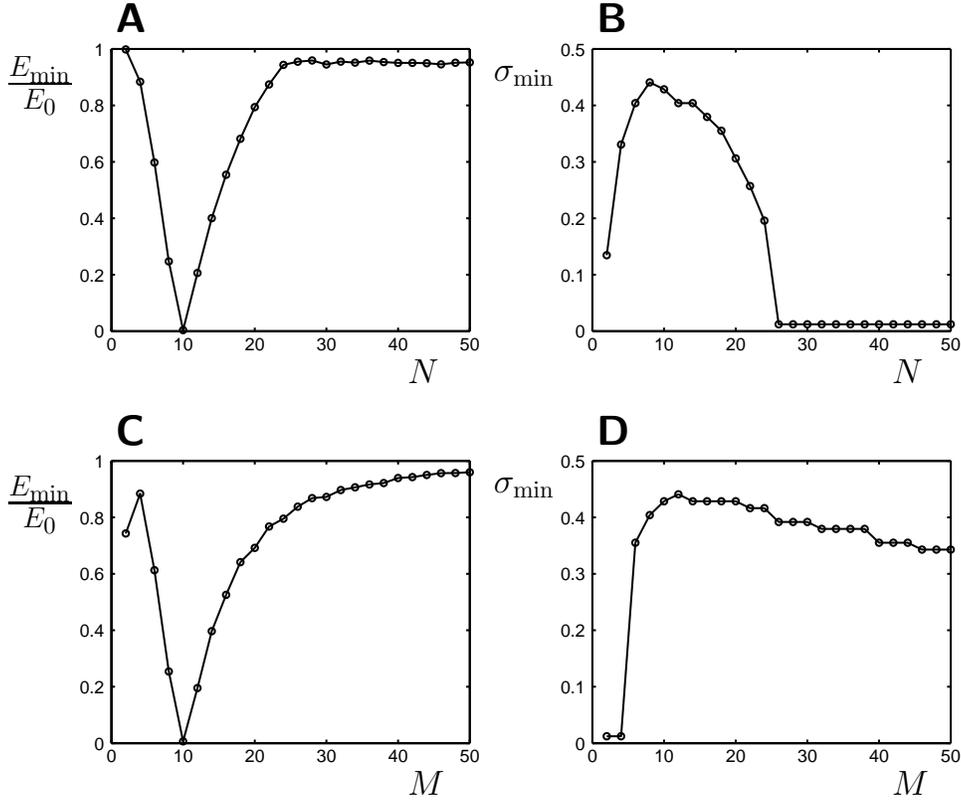,width=5.0in}}
\caption{\label{LargeNets}
Interaction between synaptic noise and response noise during the
classification of $M$ input stimuli. For each stimulus, the mean
responses of $N$ input neurons were randomly selected from a uniform
distribution between 0 and 1. The output unit of the network had to
classify the $M$ response patterns by producing either a 1 or a 0. The
synaptic noise SD was $\sigma_{W}=0.5$. Results (circles) are averages
over $1000$ networks and $100$ trials per network.  All data are from
computer simulations.
(A) Relative error, \Emin$/E_{0}$, as a function of the number of
input neurons, $N$\@.  The number of stimuli was kept constant at
$M\eq 10$.
(B) Optimal value of the response noise SD, \sigmin, as a function of
the number of input neurons, $N$\@. Same simulations as in A\@.
(C) Relative error as a function of the number of input stimuli,
$M$\@. The number of input neurons was kept constant at $N\eq 10$.
(D) Optimal value of the response noise SD as a function of $M$ for
the same simulations as in C\@. }
\end{figure*}

The results from these simulations are shown in Fig.~3. All data
points were obtained with the same amount of synaptic variability,
$\sigma_W\eq 0.5$. Each point represents an average over 1000
networks for which the optimal connections were corrupted.  The amount
of response noise that minimized the error, averaged over those 1000
corruption patterns, was found numerically by calculating the average
error with the same mean responses and corruption patterns but
different $\sigma_r$. For each combination of $N$ and $M$, this
resulted in \sigmin, which is shown in panel B\@.  The actual average
error obtained with $\sigma_r\eq$ \sigmin\ divided by the error for
$\sigma_r\eq 0$ is shown in panel A, as in the previous figure.
Interestingly, the benefit conferred by response noise depends
strongly on the difference between $N$ and $M$\@. With $M\eq 10$ input
stimuli, the effect of response noise is maximized when $N\eq 10$
neurons are used to encode them (Fig.~3A); and viceversa, when there
are $N\eq 10$ neurons in the network, the maximum effect is seen when
they encode $M\eq 10$ stimuli (Fig.~3C).  Results with other numbers
(5, 20 and 40 stimuli or neurons) were the same: response noise always
had a maximum impact when $N\eq M$\@.

This is not unreasonable. When there are many more neurons than
stimuli, a moderate amount of synaptic corruption causes only a small
error, because there is redundancy in the connectivity matrix. On the
other hand, when there are many more input stimuli than neurons, the
error is large anyway, because the $N$ neurons cannot possibly span
all the required dimensions, $M$\@. Thus, at both extremes, the impact
of synaptic noise is limited. In contrast, when $N\eq M$ there is no
redundancy but the output error can potentially be very small, so the
network is most sensitive to alterations in synaptic connectivity.
Thus, response noise makes a big difference when the number of
responses and the number of independent stimuli encoded are equal or
nearly so. In Figs.~3A, C, the relative error is not zero for
$N\eq M$, but it is quite small
(\Emin\ $\eq 0.23$, \Emin$/E_0 \eq 0.004$). This is primarily because
the error without any response noise, $E_0$, can be very large.
Interestingly, the optimal amount of response noise also seems to be
largest when $N\eq M$, as suggested by Figs.~3B, D\@.

In contrast to previous examples, for all data points in Fig.~3 the
fluctuations in the synapses and in the firing rates, $\epsilon$ and
$\eta$, were drawn from uniform rather than Gaussian distributions.
As mentioned before, the variances of the underlying distributions
should matter but their shapes should not. Indeed, with the same
variances, results for Fig.~3 were virtually identical with Gaussian
or exponential distributions.

A potential concern in this network is that, although the variability
of the output neuron depends on the interaction between the two types
of noise, perhaps the interaction is of little consequence with
respect to actual classification performance. The relevant measure for
this is the probability of correct classification, $p_c$. This
probability is obtained by comparing the distributions of output
responses to stimuli in one category versus the other, which is
typically done using standard methods from signal detection
theory~\citep{dayan-2001}. The algorithm underlying the calculation is
quite simple: in each trial, the stimulus is assumed to belong to
class 1 if the output firing rate is below a threshold, otherwise the
stimulus belongs to class 2. To obtain $p_c$, the results should be
averaged over trials and stimuli. Finally, note that an optimal
threshold should be used to obtain the highest possible $p_c$.  We
performed this analysis on the data in Fig.~3.  Indeed, $p_c$ also
depended non-monotonically on response variability.  For instance, for
$N\eq M\eq 10$ the values with and without response noise were
$p_c(\sigma_r\!= $\sigmin$)\eq 0.83$ and
$p_c(\sigma_r\eq 0)\eq 0.75$,
where chance performance corresponds to 0.5. Also, the maximum benefit
of response noise occurred for $N\eq M$ and decreased quickly as the
difference between $N$ and $M$ grew, as in Figs.~3A, C. However, the
amount of response noise that maximized $p_c$ was typically about one
third of the amount that minimized the mean square error. Thus, the
best classification probability for $N\eq M\eq 10$ was
$p_c(\sigma_r\eq 0.13)\eq 0.91$.
Maximizing $p_c$ is not equivalent to minimizing the mean square
error; the two quantities weight differently the bias and variance of
the output response (see Haykin, 1999). Nevertheless, response noise
can also counteract part of the decrease in $p_c$ due to synaptic
noise, so its beneficial impact on classification performance is real.

\section{Noise Interactions in a Sensory-Motor Network}

To illustrate the interactions between synaptic and response noise in
a more biologically realistic situation, we apply the general approach
outlined in Section~\ref{general} to a well-known model of
sensory-motor integration in the brain.  We consider the classic
coordinate transformation problem in which the location of an object,
originally specified in retinal coordinates, becomes independent of
gaze angle. This type of computation  has been thoroughly studied both
experimentally~\citep{AES85,BASG95} and
theoretically~\citep{Zipser88,Salinas+Abbott:1995,PS97}, and is
thought to be the basis for generating representations of object
location relative to the body or the world.  Also, the way in which
visual and eye-position signals are integrated here is an example of
what seems to be a general principle for combining different
information streams in the brain~\citep{ST00,Salinas+Sejnowski:2001}.
Such integration by 'gain modulation' may have wide applicability in
diverse neural circuits~\citep{Salinas-2004-2}, so it represents a
plausible and general situation in which computational accuracy is
important.

From the point of view of the phenomenon at hand, the constructive
effect of response noise, this example addresses an important issue:
whether the noise interaction is still observed when network
performance depends on a population of output neurons. In the
classification task, performance was quantified through a single
neuron's response, but in this case it depends on a nonlinear
combination of multiple firing rates, so maybe the impact of response
noise washes out in the population average. As shown below, this is
not the case.

\begin{figure*}[tb!]
\centerline{\epsfig{figure=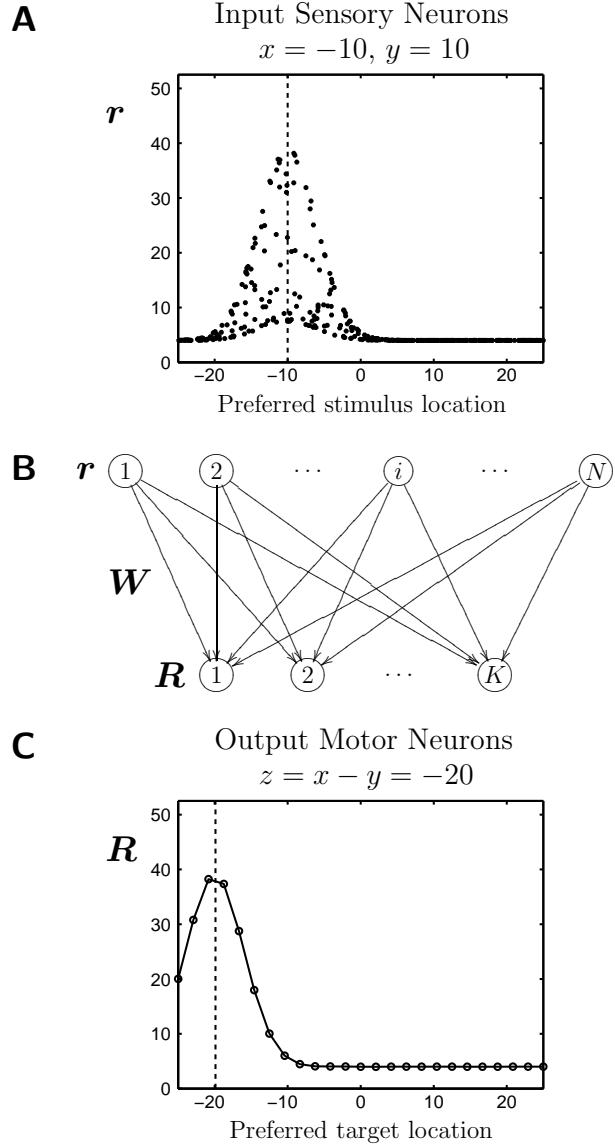,height=6.0in}}
\caption{\label{inoutSMMaps}
Network model of a sensory-motor transformation. In this network,
$N\eq 400$, $K\eq 25$, $M\eq 400$. Target and movement directions, $x$
and $z$, respectively, vary between $-25$ and $25$, whereas gaze angle
$y$ varies between $-15$ and $15$. The graphs correspond to a single
trial in which $x\eq -10$, $y\eq 10$ and $z\eq x \! - \! y\eq -20$.
Neither response noise nor synaptic corruption were included in this
example.
(A) Firing rates of the 400 gain-modulated input neurons arranged
according to preferred stimulus location.
(B) Network architecture.
(C) Firing rates of the 25 output motor neurons arranged according to
preferred target location. }
\end{figure*}

The sensory-motor network has, as before, a feedforward architecture
with two layers.  The first layer contains $N$ gain-modulated sensory
units and the second or output layer contains $K$ motor units. Each
sensory neuron is connected to all output neurons through a set of
feedforward connections, as illustrated in Fig.~4B\@. The sensory
neurons are sensitive to two quantities, the location (or direction)
of a target stimulus $x$, which is in retinal coordinates, and the
gaze (or eye-position) angle $y$.  The network is designed so that
the motor layer generates or encodes a movement in a direction $z$,
which represents the direction of the target relative to the head.
The idea is that the profile of activity of the output neurons should
have a single peak centered at direction $z$.  The correct (i.e.,
desired) relationship between inputs and outputs is $z\eq x\!-\!y$,
which is approximately how the angles $x$ and $y$ should be combined
in order to generate a head-centered representation of target
direction~\citep{Zipser88,Salinas+Abbott:1995,PS97}. In other words,
$z$ is the quantity encoded by the output neurons and it should
relate to the quantities encoded by the sensory neurons through the
function $z(x, y)\eq x\!-\!y$. Many other functions are possible, but
as far as we can tell, the choice has little impact on the qualitative
effect of response noise.

In this model, the mean firing rate of sensory neuron $i$ is
characterized by a product of two tuning functions, $f_i(x)$ and
$g_i(y)$, such that
\b
   \ol{r}_i(x, y) = r_{\mathrm{max}} \,
                   f_i(x)\left(1 - D + D\, g_i(y)\right) + r_{B} ,
   \label{rateGM}
\e
where $r_{B}\eq 4$ spikes/s is a baseline firing rate,
$r_{\mathrm{max}}\eq 35$ spikes/s and $D$ is the modulation depth,
which is set to 0.9 throughout. The sensory neurons are gain modulated
because they combine the information from their two inputs
nonlinearly. The amplitude --- but not the selectivity --- of a
visually-triggered response, represented by $f_i(x)$, depends on the
direction of gaze~\citep{AES85,BASG95,ST00}. Note that, in the
expression above, the second index of the mean rate $\ol{r}_{ij}$ has
been replaced by parentheses indicating a dependence on $x$ and $y$.
This is to simplify the notation; the responses can still be arranged
in a matrix $\ol{\bm{r}}$ if each value of the second index is
understood to indicate a particular combination of values of $x$ and
$y$. For example, if the rates were evaluated in a grid with 10 $x$
points and 10 $y$ points, the second index would run from 1 to 100,
covering all combinations. Indeed, this is how it is done in the
computer.

For simplicity, the tuning curves for different neurons in a given
layer are assumed to have the same shape but different preferred
locations or center points, which are always between $-25$ and $25$.
Visual responses are modeled as Gaussian tuning functions of stimulus
location $x$,
\b
   f_i(x) =  \exp\left(-\frac{\left(x - a_i\right)^2}{2\sigma_f^2}\right) ,
   \label{xtun}
\e
where $a_i$ is the preferred location and $\sigma_f\eq 4$ is the
tuning curve width. The dependence on eye position is modeled using
sigmoidal functions of the gaze angle $y$,
\b
   g_i(y) = \frac{1}{1 + \exp(-(b_i-y)/d_i)} \, ,
   \label{ytun}
\e
where $b_i$ is the center point of the sigmoid and $d_i$ is chosen
randomly between $-7$ and $+7$ to make sure that the curves $g_i(y)$
have different slopes for different neurons in the array.  In each
trial of the task, response variability is included by applying a
variant of Equation (\ref{inputnoise1}),
\b
   r_{ij} = \ol{r}_{ij} + \sqrt{\ol{r}_{ij}} \, \eta_{ij} .
   \label{inputnoise2}
\e
This makes the variance of the rates proportional to their means,
which in general is in good agreement with experimental data
\citep{Dean81,nc:Softky+Koch:1992,SK93,HSKD96}. This choice, however,
is not critical (see below). The desired response for each output
neuron is also described by a Gaussian,
\b
   F_k(z) = r_{\mathrm{max}} \, \exp\!\left(
                  -\frac{\left(z - c_k\right)^2}{2\sigma_F^2}
            \right) + r_{B} ,
   \label{Fout}
\e
where $\sigma_F\eq 4$ and $c_k$ is the preferred target direction of
motor neuron $k$. This expression gives the intended response of
output unit $k$ in terms of the encoded quantity $z$. Keep in mind,
however, that the desired dependence on the sensory inputs is obtained
by setting $z\eq x\!-\!y$.  When driven by the first-layer neurons,
the output rates are still calculated through a weighted sum,
\b
    R_{k}(z) = R_{k}(x, y) = \sum_{i=1}^N W_{ki} \, r_{i}(x, y) .
    \label{Rdriv1}
\e
This is equivalent to Equation (\ref{Rdriv}) but with the second index
defined implicitly through $x$ and $y$, as mentioned above. The
optimal synaptic connections $\ol{W}_{ki}$ are determined exactly as
before, using Equation~(\ref{wopt}).

Typical profiles of activity for input and output neurons are shown in
Figs.~4A, C for a trial with $x\eq -10$ and $y\eq 10$. The sensory
neurons are arranged according to their preferred stimulus location
$a_i$, whereas the motor neurons are arranged according to their
preferred movement direction $c_k$.  For this sample trial no
variability was included; the firing rate values in Fig.~4A are
scattered under a Gaussian envelope (given by Equation (\ref{xtun}))
because the gaze-dependent gain factors vary across cells. Also, the
output profile of activity is Gaussian and has a peak at the point
$z\eq -20$, which is exactly where it should be given that the correct
input-output transformation is $z\eq x\!-\!y$. With noise, the output
responses would be scattered around the Gaussian profile and the peak
would be displaced.

The error used to measure network performance is, in this case,
\b
   E_{\mathrm{pop}} = \left< \, \left| z - Z \right| \, \right> .
   \label{SMMerror}
\e
This is the absolute difference, averaged over trials and networks,
between the desired movement direction $z$ --- the actual
head-centered target direction --- and the direction $Z$ that is
encoded by the center of mass of the output activity,
\b
    Z = \frac{\sum_i \, (R_i - r_{\!B})^2 \, c_i}
             {\sum_k \, (R_k - r_{\!B})^2} \, .
    \label{centermass}
\e
Therefore, Equation (\ref{SMMerror}) gives the accuracy with which the
whole motor population represents the head-centered direction of the
target, whereas Equation (\ref{centermass}) provides the recipe to
read out such output activity.  Now the idea is to corrupt the optimal
connections and evaluate \Epop\ using various amounts of response
noise to determine whether there is an optimum.  Relative to the
previous examples, the key differences are, first, that the error in
(\ref{SMMerror}) represents a population average, and second, that
although the connections are set to minimize the average difference
between desired and driven firing rates, the performance criterion is
not based directly on it.

\begin{figure*}
\centerline{\epsfig{figure=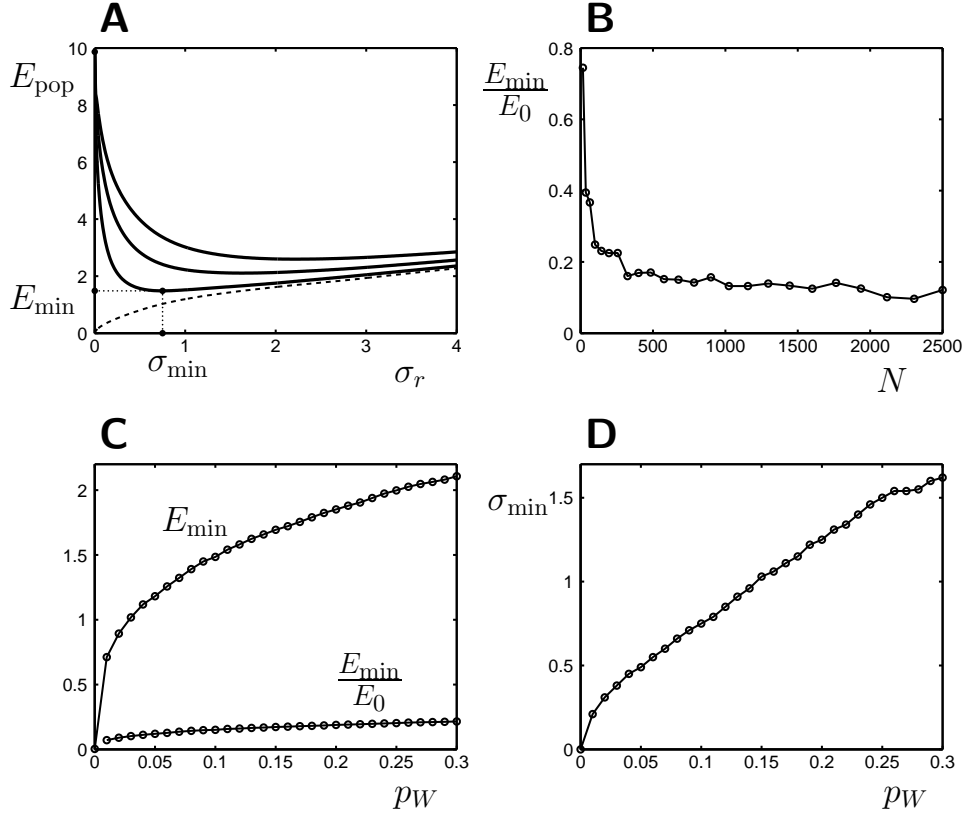,width=5.0in}}
\caption{\label{NoiseSMMaps}
Noise interaction for the sensory-motor network depicted in Fig.~4.
Results are averaged over $100$ networks and $100$ trials per network.
All data are from computer simulations.
(A) Average absolute deviation between actual and encoded target
locations, \Epop, as a function of response noise. Continuous lines
are for three probabilities of weight elimination, $p_W\eq 0.1$, 0.3
and 0.5; the dashed line corresponds to $p_W\eq 0$.
(B) Magnitude of the noise interaction, measured by the relative error
\Emin$/E_0$, as a function of the number of input neurons, $N$, for
$p_W\eq 0.2$.
(C) \Emin\ and \Emin$/E_0$ as functions of  $p_W$.
(D) Optimal response noise SD, \sigmin, as a function of $p_{W}$. }
\end{figure*}

Simulation results for this sensory-motor model are presented in
Fig.~5. A total of 400 sensory and 25 output neurons were used.  These
units were tested with all combinations of 20 values of $x$ and 20
values of $y$, uniformly spaced (thus, $M\eq 400$).  Synaptic noise
was generated by random weight elimination. This means that, after
having set the connections to their optimal values given by
Equation~(\ref{wopt}), each one was reset to zero with a probability
$p_W$. Thus, on average, a fraction $p_W$ of the weights in each
network was eliminated. As shown in Fig.~5A, when $p_W\! >\! 0$, the
error between the encoded and the true target direction has a minimum
with respect to $\sigma_r$. These error curves represent averages
over 100 networks.  Interestingly, the benefit of noise does not
decrease when more sensory units are included in the first layer
(Fig.~5B).  That is, if $p_W$ is constant, the proportion of
eliminated synapses does not change, so the error caused by synaptic
corruption cannot be reduced simply by adding more neurons.

Figure 5C shows the minimum and relative errors as functions of $p_W$.
This graph highlights the substantial impact that response noise has
on this network: the relative error stays below 0.2 even when about a
third of the synapses are eliminated. This is not only because the
error without response noise is high, but also because the error with
an optimal amount of noise stays low. For instance, with $p_W\eq 0.3$
and $\sigma_r\eq$ \sigmin, the typical deviation from the correct
target direction is about 2 units, whereas with $\sigma_r\eq 0$ the
typical deviation is about 10. Response noise thus cuts the deviation
by about a factor of five, and importantly, the resulting error is
still small relative to the range of values of $z$, which spans 50
units.  Also, as observed in the classification task, in general it is
better to include response noise even if $\sigma_r$ is not precisely
matched to the amount of synaptic variability (Fig.~5A).

Figure 5D plots \sigmin\ as a function of the probability of synaptic
elimination. The optimal amount of response noise increases with $p_W$
and reaches fairly high levels. For instance, at a value of 1, which
corresponds to $p_W$ near 0.15, the variance of the firing rates is
equal to their mean, because of Equation (\ref{inputnoise2}). We
wondered whether the scaling law of the response noise would make any
difference, so we reran the simulations with either additive noise (SD
independent of mean) or noise with an SD proportional to the mean, as
in Equation (\ref{inputnoise1}).  Results in these two cases were very
similar: \Emin\ and \Emin$/E_0$ varied very much like in Fig.~5C, and
the optimal amount of noise grew monotonically with $p_W$, as in
Fig.~5D\@.

\section{Noise Interactions in a Recurrent Network}

The networks discussed in the previous sections had a feedforward
architecture, and in those cases the contribution of response noise to
the correlation matrix between neuronal responses could be determined
analytically. In contrast, in recurrent networks the dynamics are more
complex and the effects of random fluctuations more difficult to
ascertain. To investigate whether response noise can still counteract
some of the effects of synaptic variability, we consider a recurrent
network with a well-defined function and relatively simple dynamics
characterized by attractor states. When the firing rates in this
network are initialized at arbitrary values, they eventually stop
changing, settling down at certain steady-state points in which some
neurons fire intensely and others do not. The optimal weights sought
are those that allow the network to settle at predefined sets of
steady-state responses, and the error is thus defined in terms of the
difference between the desired steady states and the observed ones. As
before, response noise is taken into account when the optimal synaptic
weights are generated, although in this case the correction it
introduces (relative to the noiseless case) is an approximation.

The attractor network consists of $N$ continuous-valued neurons, each
of which is connected to all other units via feedback synaptic
connections~\citep{hertz91b}.  With the proper connectivity, such
network can generate, without any tuned input, a steady-state profile
of activity with a cosine or Gaussian
shape~\citep{BBS95,CompteCortex00,Sali03}. Such stable `bump'-shaped
activity is observed in various neural models, including those for
cortical hypercolumns~\citep{Hansel-Sompolinsky-98}, head-direction
cells~\citep{Zhang1996,nc:laing+chow:2001} and working memory
circuits~\citep{CompteCortex00}. Below, we find the connection matrix
that allows the network to exhibit a unimodal activity profile
centered at any point within the array.

\subsection{Optimal Synaptic Weights in a Recurrent Architecture}

The dynamics of the network are determined by the equation
\b
   \tau \frac{d r_i}{d t} = -r_i
                         + h \! \left( \sum_j W_{ij} \, r_j\right)
                         + \eta_i \, ,
   \label{RNNmain}
\e
where $\tau\eq 10 $ is the integration time constant, $r_i$ is the
response of neuron $i$, and $h$ is the activation function of the
cells, which relates total current to firing rate.  The sigmoid
function
$h(x) = 1/(1 + \exp(-x))$
is used, but this choice is not critical. As before, $\eta_i$
represents the response fluctuations, which are drawn independently
for each neuron in every time step.  In this case they are Gaussian,
with zero mean and a variance $\sigma_r^2/\Delta t$. The variance of
$\eta_i$ is divided by the integration time step $\Delta t$ to
guarantee that the variance of the rate $r_i$ remains independent of
the time step~\citep{VanK01}.

For our purposes, manipulating this type of network is easier if the
equations are expressed in terms of the total input currents to the
cells~\citep{hertz91b,dayan-2001}. If the current for neuron $i$ is
$u_i \eq \sum_j W_{ij} \, r_j$, then
\b
   \tau \frac{d u_i}{d t} = -u_i + \sum_j W_{ij}
                              \left( h(u_j) + \eta_j \right) ,
   \label{RNNmain1}
\e
is equivalent to Equation (\ref{RNNmain}) above.
\begin{figure*}
\centerline{\epsfig{figure=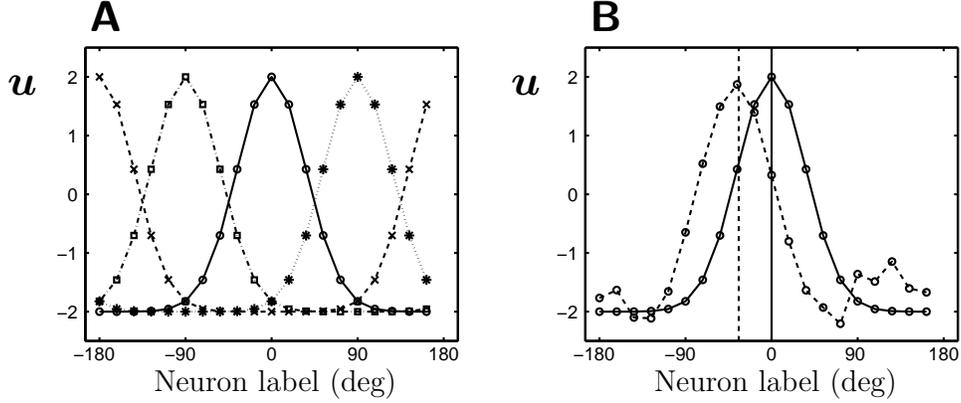,width=5.0in}}
\caption{Steady-state responses of a recurrent neural network with 20
neurons.  Results show the input currents of all units after 1000 ms
of simulation time, with responses evolving according to
Equation~(\ref{RNNmain1}). Each neuron is labeled by an angle between
-180\deg\ and 180\deg.
(A) Steady-state responses for four sets of initial conditions with
peaks near units \mbox{-90\deg}, 0\deg, +90\deg and 180\deg. The
observed activity profiles are indistinguishable from the desired
Gaussian curves.  Neither synaptic nor response noise were included in
this example.
(B) Steady-state responses with and without noise. The desired
activity profile is indicated by the solid line. The dotted line
corresponds to the activity observed with noise after 1000 ms of
simulation time, having started with an initial condition equal to the
desired steady state. Vertical lines indicate the locations of the
corresponding centers of mass. The absolute deviation is 34\deg.
Here, $\sigma_r\eq 0.3$ and $p_W\eq 0.02$. }
\end{figure*}
A stationary solution of Equation (\ref{RNNmain1}) without input noise
is such that all derivatives become zero. This corresponds to an
attractor state $\alpha$ for which
\b
   u_i^{\alpha} = \sum_j W_{ij} \, h(u_j^{\alpha}) .
   \label{SScondition}
\e
The label $\alpha$ is used because the network may have several
attractors or sets of fixed points. The desired steady-state currents
are denoted as $U_i^{\alpha}$. These are Gaussian profiles of activity
such that, during steady state $\alpha\eq 1$, neuron 1 is the most
active (i.e., the Gaussian is centered at neuron 1), during steady
state $\alpha\eq 2$, neuron 2 is the most active, and so on. Figure 6
illustrates the activity of the network at four steady states in the
absence of noise ($\sigma_W\eq 0\eq \sigma_r$). To make the network
symmetric, the neurons were arranged in a ring, so their activity
profiles wrap around. Because of this, each neuron is labeled with an
angle. The observed currents $u_i$ settle down at values that are
almost exactly equal to the desired ones, $U_i^{\alpha}$. The synaptic
connections that achieved this match were found by enforcing the
steady-state condition (\ref{SScondition}) for the desired attractors.
That is, we minimized
\b
   E = \frac{1}{N_A} \sum_{\alpha = 1}^{N_A} \sum_{i} \left(
               U_i^{\alpha} - \sum_j W_{ij} \, h(U_j^{\alpha})
       \right)^{\!2} ,
   \label{RNNerror}
\e
where $U_i^{\alpha}$ is a (wrap-around) Gaussian function of $i$
centered at $\alpha$ and $N_A$ is the number of attractors; in the
simulations $N_A$ is always equal to the number of neurons, $N$\@.
This procedure leads to an expression for the optimal weights
equivalent to Equation (\ref{wopt}). Thus, without response noise,
\b
    \ol{\bm{W}} = \bm{L} \, \bm{C}^{-1} ,
    \label{RNNwopt}
\e
where
\ba
    L_{ij} & = & \frac{1}{N_A} \sum_{\alpha}
                 U_i^{\alpha} \, h(U_j^{\alpha}) \nonumber \\
    C_{ij} & = & \frac{1}{N_A} \sum_{\alpha}
                 h(U_i^{\alpha}) \, h(U_j^{\alpha}) \, .
    \label{RNNcorr}
\ea
To include the effects of response noise, we add a correction to the
diagonal of the correlation matrix, as in the previous cases (see
Section \ref{RegSect}). We thus set
\b
    C_{ij} = \frac{1}{N_A} \sum_{\alpha}
               h(U_i^{\alpha}) h(U_j^{\alpha})
               + \delta_{ij} \, a \, \frac{\sigma_r^2}{2 \tau} ,
    \label{RNNapprox}
\e
where $a$ is a proportionality constant. The rationale for this is as
follows.

Strictly speaking, Equation (\ref{RNNmain1}) with response noise does
not have a steady state. But consider the simpler case of a single
variable $u$ with a constant asymptotic value $u_{\infty}$, such that
\b
    \tau \frac{d u}{d t} = -u + u_{\infty} + \eta .
    \label{singleu}
\e
If the trajectory $u(t)$ from $t\eq 0$ to $t\eq T$ is calculated many
times, starting from the same initial condition, the distribution of
endpoints $u(T)$ has a well-defined mean and variance, which vary
smoothly as functions of $T$\@. The mean is always equal to the
endpoint that would be observed without noise, whereas for $T$ much
longer than the integration time constant $\tau$, the variance is
equal to the variance of the fluctuations on the right hand side of
Equation (\ref{singleu}) divided by $2\tau$~\citep{VanK01}. These
considerations suggest that we minimize
\b
   E = \frac{1}{N_A} \sum_{\alpha,i} \left(
              U_i^{\alpha} - \sum_j W_{ij} \,
                  \left( h(U_j^{\alpha}) + a \, \tilde{\eta}_j \right)
       \right)^{\!2} ,
   \label{RNNerror1}
\e
where the variance of $\tilde{\eta}_j$ is $\sigma_r^2/(2\tau)$. This
leads to Equation (\ref{RNNwopt}) with the corrected correlation
matrix given by (\ref{RNNapprox}).

\subsection{Performance of the Attractor Network}

To evaluate the performance of this network, we compare the center of
mass of the desired activity profile to that of the observed profile
tracked during a period of time. For a particular attractor $\alpha$,
the network is first initialized very close to that desired steady
state, then Equation (\ref{RNNmain1}) is run for 1000 ms (100 time
constants $\tau$), and the absolute difference between the initial and
the current centers of mass is recorded during the last 500 ms.  The
error for the recurrent networks \Erec\ is defined as the absolute
difference averaged over this time period and all attractor states,
ie., all values of $\alpha$.  Also, when there is synaptic noise, an
additional average over networks is performed.  This error function is
similar to Equation (\ref{SMMerror}), except that the circular
topology is taken into account.  Thus, \Erec\ is the mean absolute
difference between desired and observed centers of mass.  It is
expressed in degrees.

\begin{figure*}[t!]
\centerline{\epsfig{figure=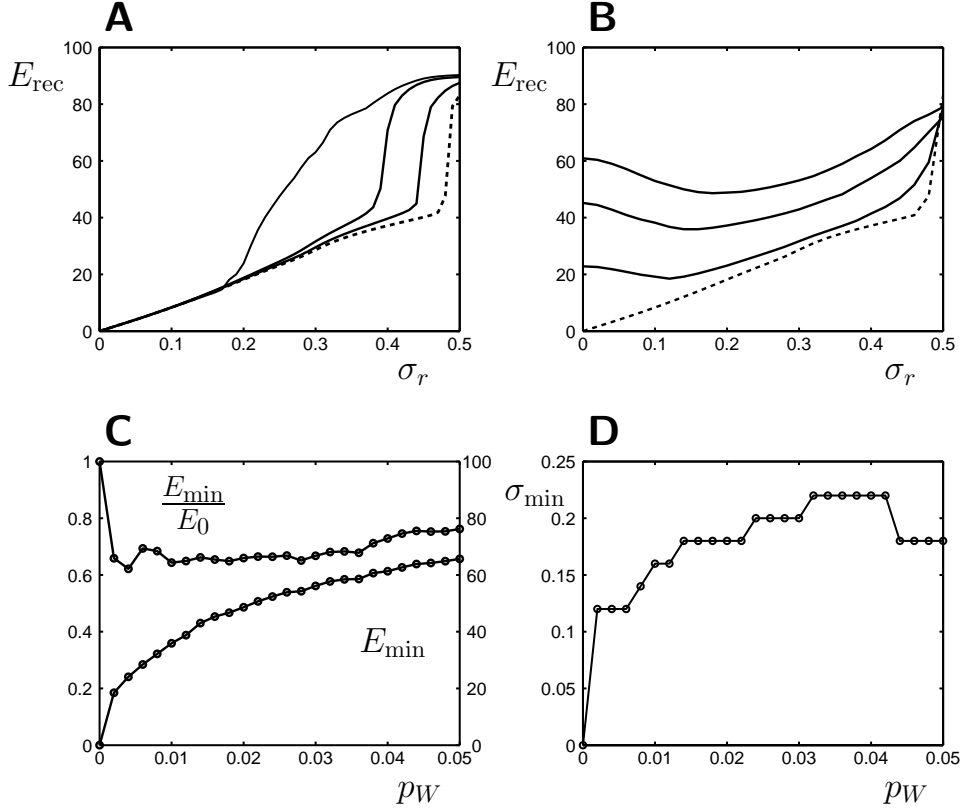,width=5.0in}}
\caption{Interaction between synaptic and response noise in
recurrent networks. (A) Average absolute difference between desired
and observed centers of mass as a function of $\sigma_r$. Units are
degrees. The different curves are for $a\eq 0$, 1.5, 1 and 0.5, from
left to right. The lowest curve (dashed) was obtained with $a\eq
0.5$, confirming that the synaptic weights are optimized when
response noise is taken into account. (B) Average error \Erec\ as a
function of response noise. Continuous lines are for three
probabilities of weight elimination $p_W\eq 0.005$, 0.015 and 0.025;
the dashed line corresponds to $p_W\eq 0$.  Here and in the
following panels, $a\eq 0.5$. (C) \Emin$/E_0$ (left y-axis) and
\Emin\ (right y-axis) as functions of $p_W$. (D) Optimal response
noise SD, \sigmin, as a function of $p_{W}$ for the same data in C.
}
\end{figure*}

Before exploring the interaction between synaptic and response
noise, we used \Erec\ to test whether the noise-dependent correction
to the correlation matrix in Equation (\ref{RNNapprox}) was
appropriate. To do this, a recurrent network without synaptic
fluctuations was simulated multiple times with different values of
the parameter $a$ and various amounts of response noise. The desired
attractors were kept constant.  The resulting error curves are shown
in Fig.~7A\@. Each one gives the average absolute deviation between
desired and observed centers of mass as a function of $\sigma_r$ for
a different value of $a$. The dependence on $a$ was non-monotonic.
The optimal value we found was 0.5, which corresponds to the lowest
curve (dashed) in the figure. This curve was well below the one
observed without adjusting the synaptic weights.  Therefore, the
correction was indeed effective.

Figure 7B shows \Erec\ as a function of $\sigma_r$ when synaptic
noise is also present in the recurrent network. The three solid
curves correspond to nets in which synapses were randomly eliminated
with probabilities $p_W\eq 0.005$, 0.015 and 0.025. As with previous
network architectures, a non-zero amount of response noise improves
performance relative to the case where no response noise is
injected. In this case, however,  the mean absolute error is already
about 25\deg at the point at which response noise starts making a
difference, around $p_W\eq 0.005$ (Fig.\ 7C). This is not
surprising: these types of networks are highly sensitive to changes
in their synapses, so even small mismatches can lead to large
errors~\citep{SLRT00,RSW03}.  Also, Fig.~7C shows that the ratio
\Emin$/E_0$ does not fall below 0.6, so the benefit of noise is not
as large as in previous examples. The effect was somewhat weaker
when synaptic variability was simulated using Gaussian noise with SD
$\sigma_W$ instead of random synaptic elimination. Nevertheless, it
is interesting that the interaction between synaptic and response
noise is observed at all under these conditions, given that the
response dynamics are richer and that the minimization of Equation
(\ref{RNNerror1}) may not be the best way to produce the desired
steady-state activity.

\section{Discussion}

\subsection{Why are Synaptic and Response Fluctuations Equivalent?}
\label{disc1}

We have investigated the simultaneous action of synaptic and response
fluctuations on the performance of neural networks and found an
interaction or equivalence between them: when synaptic noise is
multiplicative, its effect is similar to that of response noise. At
heart, this is a simple consequence of the product of responses and
synaptic weights contained in most neural models, which has the form
$\sum_j W_j r_j$. With multiplicative noise in one of the variables,
this weighted sum turns into $\sum_j W_j (1 + \xi_j) r_j$, which is
the same whether it is the synapse or the response that fluctuates. In
either case, the total stochastic component $\sum_j W_j \xi_j r_j$
scales with the synaptic weights. The same result is obtained with
additive response noise.  Additive synaptic noise behaves differently,
however. It instead leads to a total fluctuation $\sum_j \xi_j r_j$
that is independent of the mean weights.  Evidently, in this case the
mean values of the weights have no effect on the size of the
fluctuations.  Thus, the key requirement for some form of equivalence
between the two noise sources is that the synaptic fluctuations must
depend on the strength of the synapses.

This condition was applied to the three sets of simulations presented
above, which corresponded to the classification of arbitrary response
patterns, a sensory-motor transformation, and the generation of
multiple self-sustained activity profiles. This selection of problems
was meant to illustrate the generality of the observations outlined in
the above paragraph. And indeed, although the three problems differed
in many respects, the results were qualitatively the same.

We should also point out that, in all the simulations, the criterion
used to determine the optimality of the synaptic weights was based on
a mean square error. But perhaps the noise interaction changes when a
different criterion is used. To investigate this, we performed
additional simulations of the small $2\! \times\! 1$ network in which
the optimal synaptic weights were those that minimized a mean absolute
deviation; thus, the square in Equation (\ref{error}) was substituted
with an absolute value. In this case everything proceeded as before,
except that the mean weight values $\ol{W}$ had to be found
numerically. For this, the averages were performed explicitly and the
downhill simplex method was used to search for the best
weights~\citep{PFTV92}. The results, however, were very similar to
those in Fig.~2A\@.  Although the shapes of the curves were not
exactly the same, the relative and minimum errors found with the
absolute value varied very much like with the mean-square error
criterion as functions of $\sigma_W$. Therefore, our conclusions do
not seem to depend strongly on the specific function used to weight
the errors and find the best synaptic connection values.

\subsection{When Should Response Noise Increase?}
\label{disc2}

According to the argument above, the most general way to state our
results is this: assuming that neuronal activities are determined by
weighted sums, any mechanism that is able to dampen the impact of
response noise will automatically reduce the impact of multiplicative
synaptic noise as well. Furthermore, we suggest that under some
circumstances it is better to add more response noise and increase the
dampening factor, than ignore the synaptic fluctuations altogether.
There are two conditions for this scenario to make sense.  (1) The
network must be highly sensitive to changes in connectivity.  This can
be seen, for instance, in Fig.~3A, which shows that the highest
benefit of response noise occurs when the number of neurons matches
the number of conditions to be satisfied --- it is at this point that
the connections need to be most accurate.  (2) The fluctuations in
connectivity cannot be evaluated directly.  That is, why not take into
account the synaptic noise in exactly the same way as the response
noise when the optimal connections are sought?  For example, the
average in Equation (\ref{errormatrix}) could also include an average
over networks (synaptic fluctuations), in which case the optimal mean
weights would depend not only on $\sigma_r$ but also on $\sigma_W$. In
the simulations this could certainly be done, and would lead to
smaller errors. But we explicitly consider the possibility that either
$\sigma_W$ is unknown a priori, or there is no separate biophysical
mechanism for implementing the corresponding corrections to the
synaptic connections.

Condition number 2 is not unreasonable. Realistic networks with high
synaptic plasticity must incorporate mechanisms to ensure that ongoing
learning does not disrupt their previously acquired functionality.
Thus, synaptic modifications rules need to achieve two goals: to
establish new associations that are relevant for the current
behavioral task, and to make adjustments to prevent interference from
other, future associations. The latter may be particularly difficult
to achieve if learning rates change unpredictably with time.  It is
not clear whether plausible (e.g., local) synaptic modification
mechanisms could solve both problems simultaneously (see Hopfield and
Brody, 2004), but the present results suggest an alternative: synaptic
modification rules could be used exclusively to learn new associations
based on current information, whereas response noise could be used to
indirectly make the connectivity more robust to synaptic fluctuations.
Although this mechanism evidently doesn't solve the problem of
combining multiple learned associations, it might alleviate it. Its
advantage is that, assuming that neural circuits have evolved to
adaptively optimize their function in the face of true noise, simply
increasing their response variability would generate synaptic
connectivity patterns that are more resistant to fluctuations.

\subsection{When is Synaptic Noise Multiplicative?}
\label{disc3}

The condition that noise should be multiplicative means that changes
in synaptic weight should be proportional to the magnitude of the
weight.  Evidently, not all types of synaptic modification processes
lead to fluctuations that can be statistically modeled as
multiplicative noise; for instance, saturation may prevent positive
increases, thus restricting the variability of strong synapses.
However, synaptic changes that generally increase with initial
strength should be reasonably well approximated by the multiplicative
model.  Random synapse elimination fits this model because, if a weak
synapse disappears, the change is small, whereas if a strong synapse
disappears, the change is large. Thus, the magnitude of the changes
correlates with initial strength. Another procedure that corresponds
to multiplicative synaptic noise is this.  Suppose the size of the
synaptic changes is fixed, so that weights can only vary by
$\pm \delta w$, but suppose also that the probability of suffering a
change increases with initial synaptic strength. In this case, all
changes are equal, but on average a population of strong synapses
whould show higher variability than a population of weak ones. In
simulations, the disruption caused by this type of synaptic corruption
is indeed lessened by response noise (data not shown).

\subsection{Final Remarks}
\label{disc4}

To summarize, the scenario we envision rests on five critical
assumptions: (1) the activity of each neuron depends on
synaptically-weighted sums of its (noisy) inputs, (2) network
performance is highly sensitive to changes in synaptic connectivity,
(3) synaptic changes unrelated to a function that has already been
learned can be modeled as multiplicative noise, (4) synaptic
modification mechanisms are able to take into account response noise,
so synaptic strengths are adjusted to minimize its impact, but (5)
synaptic modification mechanisms do not directly account for future
learning. Under these conditions, our results suggest that increasing
the variability of neuronal responses would, on average, result in
more accurate performance. Although some of these assumptions may be
rather restrictive, the diversity of synaptic plasticity mechanisms
together with the high response variability observed in many areas of
the brain make this constructive noise effect worth considering.

\subsubsection*{Acknowledgments.}
Research was supported by NIH grant NS044894.

%\bibliography{neuroscience}

\end{document}